\PassOptionsToPackage{table,xcdraw,svgnames}{xcolor}
\PassOptionsToPackage{colorlinks,pagebackref,bookmarks}{hyperref}

\documentclass[preprint,journal]{vgtc}                 

\graphicspath{{figures/}{pictures/}{images/}{./}} 

\usepackage{times}                     

\usepackage{tabu}                      
\usepackage{booktabs}                  
\usepackage{lipsum}                    
\usepackage{mwe}                       

\usepackage{mathptmx}                  

\usepackage{amssymb}
\usepackage{amsmath}
\usepackage{multirow}
\usepackage{tcolorbox}
\tcbuselibrary{breakable}
\usepackage{soul}
\usepackage{array}
\usepackage{xspace}
\usepackage{caption}
\usepackage{multirow}
\usepackage{ragged2e}
\usepackage{balance}
\usepackage{enumitem}
\usepackage{subcaption}
\usepackage{float}
\usepackage{stfloats}
\usepackage{microtype}

\usepackage{hyperref}

\onlineid{1012}

\vgtccategory{Research}

\vgtcinsertpkg

\preprinttext{To appear in the 1st Workshop on Accessible Data Visualization at IEEE VIS 2024.}

\newcommand{\iDotE}{i.\,e.,\xspace}
\newcommand{\eDotG}{e.\,g.,\xspace}

\title{Toward Understanding the Experiences of People in Late Adulthood with Embedded Information Displays in the Home}

\author{Zack While\thanks{e-mail: zwhile@cs.umass.edu}\\ 
        \scriptsize UMass Amherst %
\and Henry Wheeler-Klainberg\thanks{e-mail: hwheelerklai@umass.edu}\\ %
     \scriptsize UMass Amherst %
\and Tanja Blascheck\thanks{e-mail: research@blascheck.eu}\\ %
     \scriptsize University of Stuttgart%
\and Petra Isenberg\thanks{e-mail: petra.isenberg@inria.fr}\\ %
     \parbox{1.4in}{\scriptsize \centering Universit\'e Paris-Saclay, CNRS, Inria, LISN}
\and Ali Sarvghad\thanks{e-mail: asarv@cs.umass.edu}\\ %
     \scriptsize UMass Amherst %
     }

\abstract{
\textit{Embedded information displays} (EIDs) are becoming increasingly ubiquitous on home appliances and devices such as microwaves, coffee machines, fridges, or digital thermostats. These displays are often multi-purpose, functioning as interfaces for selecting device settings, communicating operating status using simple visualizations, and displaying notifications. However, their usability for \textit{people in the late adulthood} (PLA) development stage is not well-understood. 
We report on two focus groups with PLA ($n=11$, ages 76--94) from a local retirement community. Participants were shown images of everyday home electronics and appliances, answering questions about their experiences using the EIDs. 
Using open coding, we qualitatively analyzed their comments to distill key themes regarding how EIDs can negatively affect PLA's ability to take in information (\eDotG poor labels) and interact with these devices (\eDotG unintuitive steps) alongside strategies employed to work around these issues. 
We argue that understanding the equitable design and communication of devices' functions, operating status, and messages is important for future information display designers. We hope this work stimulates further investigation into more equitable EID design.
} 


\begin{document}


\firstsection{Introduction}

\maketitle

\firstsection{Introduction}\label{sec:introduction}
\textit{Embedded information displays} (EIDs) are interfaces that show information pertinent to a device's operations. They are increasingly ubiquitous in modern homes and connected to various devices such as ovens, washing machines, and thermostats. These displays use a variety of technologies for representing information, ranging from single LED lights to complex LCD screens.  
EIDs can also introduce various usability problems for \textit{people in the late adulthood} (PLA) development stage~\cite{berk2022development} due to changes with aging such as low color contrast~\cite{davin2022home}, small font sizes~\cite{tsuchiya2018study}, and difficulty finding or understanding information in complex interfaces~\cite{ghorayeb2023development}---corresponding to similar results in visualization work regarding low contrast~\cite{morey2019mobile}, small font sizes~\cite{fan2023understanding}, and difficulty with understanding visually complex representations~\cite{le2015evaluation}. 
However, the experiences of PLA when reading and understanding information from EIDs are poorly understood.
Most existing work related to PLA and information communication on home devices instead centers on smart homes (\eDotG Chang and Östlund~\cite{chang2019perspectives}), leaving out many devices that PLA may interact with daily.
Furthermore, recent work has advocated for greater focus on PLA~\cite{while2024gerontovis} alongside broader calls for reaching wider audiences with data visualization~\cite{lee2020reaching}.
This work begins filling an important gap in understanding how EIDs communicate information to PLA.

We report our investigation into the experiences of PLA in using home electronics with EIDs. 
We conducted two focus groups at a retirement community, engaging participants in discussion about the EIDs on the home devices available at the facility and having them share their daily experiences interacting with them.
Through a thematic analysis of their comments, we observed that most negative experiences centered on reading information from these devices, difficulties interacting with them, and strategies for working around these challenges. We contribute first empirical evidence of extant accessibility issues in these devices for PLA, as well as preliminary insights into key areas for future design improvement.

Lastly, as a part of the \textit{AccessViz} workshop, it is important to clarify how EIDs and aging relate to visualization and accessibility, respectively. While EIDs may not use traditional visualizations such as bar graphs, they visually encode and display information critical to a device's usage. The type of information (\eDotG temperature) and encoding (\eDotG lights) varies by device, however both EIDs and visualizations fall under a larger umbrella of \textit{visual information communication}.
Aging, on the other hand, has a more complex relationship with accessibility. While recent work in HCI cautions against conflating aging with accessibility when designing technology in the context of PLA~\cite{knowles2021harm}, our work focuses on how information interface design interacts with perception and cognition in the context of PLA. Aging has a broad correlation with progressive changes in these two aspects~\cite{while2024gerontovis}, potentially leading to inequities in EID and visualization usability and thus being of interest to this workshop and the visualization community. However, While et al.~\cite{while2024gerontovis} argue that accessibility research on its own does not fully suffice for supporting PLA due to the group's heterogeneity, possible compounding effects of multiple changes in perception and cognition due to aging, and differences in expectations for technology compared to younger participants in accessibility studies.
Ultimately, further work and discussion are needed in order to properly define the visualization community's approach toward aging-related work as part of the broader HCI community. We hope this work can serve as a bridge between aging and accessibility, sparking conversations about their relationship in the context of visualization and EIDs.

\section{Related Work}\label{sec:related-work}

Here we discuss the design of EIDs, designs of EIDs specifically with PLA in mind, and barriers to technology adoption for PLA. 
Existing work predominantly addresses other aspects of EID design, leaving a significant gap in our understanding of how to facilitate effective and accessible information communication for PLA.

\textbf{\textit{EID Design.}}
Existing work in EID design has had several areas of primary focus. Sepahpour et al.~\cite{sepahpour2021connection}, for example, found that device \textit{functionality} (\eDotG usability and performance) led to the strongest positive and negative emotional reactions compared to \textit{aesthetic} (\eDotG form and color) and \textit{symbolic} (\eDotG brand and memories) attributes. 
Bartram et al.~\cite{bartram2015design} recognized \textit{ecological} (\iDotE how well the device's size and location suits its environment) and \textit{aesthetic} (\iDotE visual appeal) fit for home eco-feedback displays, while Wood et al.~\cite{wood2007influencing} noted that home energy displays should be close to the device they monitor and should avoid overwhelming the user with information.
These works highlight important high-level considerations but neither illustrate the design choices that can create usability issues in understanding EIDs nor what those issues are.

\textbf{\textit{EID Design for PLA}}. 
Some work has looked at the design of embedded information displays with PLA in mind, giving recommendations to increase usability.  
Zhao et al.~\cite{zhao2024research} found that exercise bike interfaces for PLA should prioritize intuitiveness and visibility, readable information, and a sensible button layout that visually distinguishes heavily-used functions. 
This echoes other work in broader device design for PLA that encourages visually distinct color for display icons~\cite{consolvo2004carenet, owsley2011older}, and larger text~\cite{owsley2011older}. Intuitive design is critical because too many functions or poorly labeled buttons can be overwhelming for PLA~\cite{mitzner2010older,vaportzis2017older}.  
Button size and spacing are also crucial, with Jin et al.~\cite{jin2007touch} recommending specific button dimensions based on users' manual dexterity and reaction times while cautioning against offering too little (low accuracy) or too much (high search time) space between buttons. Additionally, they acknowledged the trade-offs of size and spacing when screen real estate is limited.
While these papers focus on the usability of screens and touch interfaces for PLA, our work focuses on how EIDs communicate information to PLA, which can take multiple forms including screens, lights, and labelled buttons.

\textbf{\textit{PLA's Barriers to Technology Adoption.}}
PLA may encounter a variety of obstacles both when deciding to adopt technology and after acceptance. 
Lee et al.~\cite{lee2015perspective} and Yusif et al.~\cite{yusif2016older} each reviewed papers associated with barriers for PLA's willingness to adopt technology. The former emphasized that PLA are often hesitant to use a piece of technology unless it has both perceived and tangible benefits, while the latter observed that assistive technology often needs to address a specific need. Convenience is also a noteworthy adoption consideration for PLA~\cite{mitzner2010older,vaportzis2017older}.
Challenges can also arise after acceptance of a device, including difficulty with understanding manuals~\cite{owsley2011older} and a lack of instructions for complex devices such as tablets~\cite{vaportzis2017older}; to reduce these barriers, increasing ease of use and offering technical support are recommended to facilitate continued use~\cite{lee2015perspective}. Emotional barriers such as fear of dependency, social isolation, and feelings of inadequacy also impact how they view technology~\cite{lee2015perspective, vaportzis2017older}. We expand on these by also considering how EID designs can limit \textit{how} PLA use a device, as many devices in the study were already adopted by participants but had limited usage. 

\section{Study Methodology}\label{sec:study}
Our study employed a focus group methodology to collect PLA's experiences using EIDs. We collected data through audio recordings and qualitatively analyzed it, as reported next.

\subsection{Study Design}
Our goal in running this study was to engage directly with PLA in a discussion about EIDs on devices they are familiar with. Thus, we designed the study as two focus groups with PLA at a retirement community.
Having multiple PLA in a focus group brought diverse perspectives, encouraged dynamic interactions between participants, and fostered consensus, increasing the richness of our data. Running the study at a retirement community also provided a population with a high likelihood of having similar or the same basic appliances in their homes, making discussion easier. We fine-tuned study questions and stimuli (images of devices) using a pilot study. 

\subsection{Participants}
After IRB approval, we contacted a local retirement community and got approval to run the study there and recruit residents to participate. We sent digital and printed flyers advertising the study, gathering 11 participants (1 male, 10 female) with ages ranging from 76 to 94 (\textit{Avg. = $87$}). Education levels were high school diploma ($n=1$), bachelor's degree ($n=2$), master's degree ($n=7$), and doctorate ($n=1$).
Participants had resided there for 1--5 years ($n=8$), 5--10 years ($n=2$), and more than 10 years ($n=1$). Most participants ($n=9$) had worked white-collar jobs before retiring, compared to the service industry ($n=1$) and blue-collar jobs ($n=1$). All participants self-reported their familiarity with traditional data visualizations (\eDotG bar charts, pie charts, and scatterplots) on a 5-point Likert scale~\cite{likert1932technique} as a 3 or higher, with 4/5 as the most common rating ($n=7$) compared to 3/5 ($n=2$) and 5/5 ($n=2$).

With these participants, we ran two focus groups, with five and six participants, respectively; each lasted approximately one hour, and participants were paid a \$25 Amazon gift card for their time.

\subsection{Stimuli}
\label{sec:stimuli}
\begin{figure*}
\centering
\includegraphics[width=\linewidth]{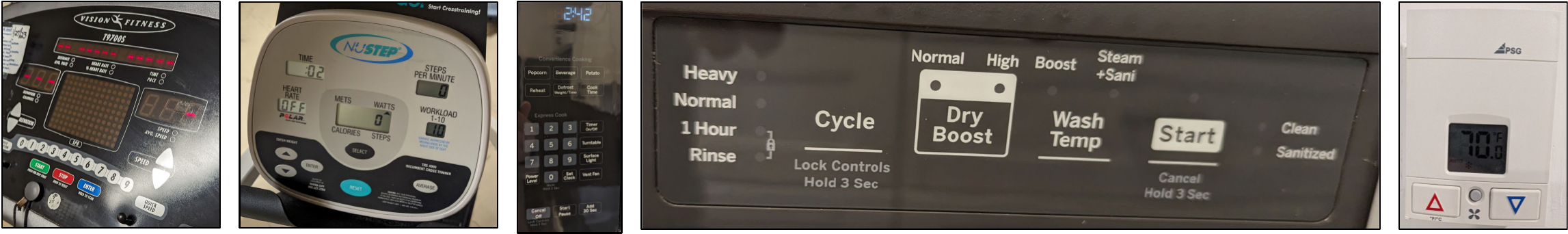}
\caption{Some of the EIDs shown to participants: Treadmill, Stepping Machine, Microwave, Dishwasher, and Thermostat.} 
\label{fig:devices}
\end{figure*}

To understand which devices would be available to and used by participants, we first visited the retirement community. We toured an example apartment and common areas, inquiring about the devices residents typically used. This resulted in a final set of 10 devices, from both common areas (microwave, two different stepping machines, treadmill, and digital community information board) and individual apartments (washing machine, dryer, microwave, stove, and thermostat), with some examples shown in \autoref{fig:devices}. During the focus group, participants were shown slides with images of these devices and detailed design schematics illustrating various functionalities and types of information available on the EIDs.

\subsection{Study Procedure}
Participants first filled out a consent form and demographic questionnaire, followed by an explanation of the study's goals and format. During each focus group, participants were shown stimuli (described in further detail in \autoref{sec:stimuli}) one at a time. Participants were asked about their experiences with the EIDs and any difficulties using them, with follow-up questions based on their responses.

\subsection{Data Collection and Analysis}
Participants' feedback was audio-recorded and later transcribed for qualitative analysis. We used \textit{inductive coding}~\cite{chandra2019inductive}, allowing for themes to emerge from the data. We first had 4 authors read the transcripts and generate their own sets of codes using open coding~\cite{vollstedt2019introduction}. These codes were then merged and restructured by 2 of the authors using axial coding~\cite{vollstedt2019introduction}, resulting in a final set of codes.
Two of the authors then re-applied this merged set of codes to the transcripts of each study, which we then used for thematic analysis.

\section{Data Analysis and Results}\label{sec:results}

The results of our thematic analysis are organized into three major themes: challenges with \textit{reading} information on EIDs, difficulties \textit{interacting} with embedded information displays, and \textit{compensating strategies} for alleviating some accessibility issues.

\subsection{What Obstacles Exist for PLA When Reading Information on EIDs?}
Throughout the focus groups, participants mentioned difficulty using the devices due to difficulty reading or understanding EIDs. This can result in a device having limited or no usability.

\textbf{\textit{Lack of Information Clarity.}}
Some participants commented on ambiguities in the information presented on the EIDs. For example, when discussing the labeled buttons on their dishwasher, \textbf{P11} mentioned that \textit{``I don't understand what those words mean, you know? I mean when it says dry boost what is it doing?''}, with other participants agreeing that the function label (and others such as the \textsc{Normal} setting) lacked a clear indication of what they entailed. Comments such as this may indicate a gap in assumptions for the design language of more modern devices, requiring users to read a manual or assume its functionality due to the lack of context.

\textbf{\textit{Complicated or Overwhelming Presentation.}}
The amount of information on the EIDs as well as how it was presented was another topic of discussion. \textbf{P8} noted that the exercise metrics such as heart rate and distance on one stepping machine were \textit{``easy to ignore''} when using the device because \textit{``they're too complicated to be of good use.''} \textbf{P4} described the EID on their thermostat as \textit{``a mess''} due to the amount of small icons and labels, with \textbf{P3} adding that they \textit{``don't know what [the icons and labels] mean.''} 

\textbf{\textit{Poor Perceptibility.}}
Several participants commented on their difficulty reading printed labels and icons on various devices. The dishwasher sparked a lot of discussion, with \textbf{P6} mentioning that the text is \textit{``too gray, it's too much...I think it might be so faint that I don't even see it on my screen.''} with \textbf{P7} immediately stating, \textit{``Yeah, I've never used it because of that.''} Comments attributed these issues to factors such as the font color, the viewing angle (below the counter), and the lack of light, with \textbf{P3} saying that the text is \textit{``not in any kind of, you know, illuminated text or anything that really shows up''}. 
Participants also commented on the size of text and icons on EIDs, such as \textbf{P1}, mentioning that the thermostat icons are \textit{``so tiny you can't see them.''} 
Screen size was similarly brought up as an issue, with \textbf{P6} describing how \textit{``I don't have any trouble with all the symbols in this case, but the whole screen is too small. And also in our case, it's too high up, in the room it's too high.''}
Auditory signals also came up, with \textbf{P3} and \textbf{P6} mentioning difficulty hearing alerts from their microwave and stove, respectively; both \textbf{P1} and \textbf{P7} wished their washing machines had sound alerts to let them know they finished running without being in the same room. 

\textbf{\textit{Difficulty Finding Functionality.}}
Some participants talked about the lack of consistency in EID design, with \textbf{P1} and \textbf{P5} mentioning that microwaves and thermostats, respectively, are \textit{``all different.''} \textbf{P4} agreed, stating that \textit{``When I'm made queen of the universe, I'm going to have every microwave be the same. Why don't they do that?''} \textbf{P3} mentioned issues finding the buttons for the timer (counting down for a specified amount of time) and time cook (microwaving something for a specified amount of time) on the microwave, so as a result \textit{``I constantly have to make sure whether I'm pushing the timer or the time I want it to cook}.'' Many participants noted that they were unaware of EID functionalities mentioned by others during the focus group, including those from the microwave (\textbf{P2}, \textbf{P7}, and \textbf{P11}), thermostat (\textbf{P2} and \textbf{P4}), and treadmill (\textbf{P1}). 

\subsection{How can EIDs Affect PLA's Ability to Interact?}
EIDs play an important role in informing and facilitating interactions with electronic devices. As a result, poor EID design can have a negative impact on a device's usability.

\textbf{\textit{Difficulty Discovering Interaction.}}
Participants commented on their experiences learning how to utilize EIDs as gateways to access and use device functions.
\textbf{P4} said that learning how to use their dishwasher was \textit{``not intuitive...I had to get maintenance to come and teach me.''} 
Only one participant (\textbf{P6}) mentioned using a manual to learn to interact with the EID; however, that approach was \textit{``much too slow for me to find out, and I can't remember it.''} Two other participants (\textbf{P1} and \textbf{P3}) mentioned difficulties remembering steps to use devices, with both referring to one of the public stepping machines. Overall, \textbf{P6} summarized their experiences as \textit{``when you're moving to an apartment, it might take you some time to figure out how to use each device.''} 

\textbf{\textit{Poor Interaction Support.}}
If an EID does not provide feedback or helpful instructions, necessary interactions may be unclear.
\textbf{P6} noted that button sequences for their stovetop's additional functions comprised of \textit{``different combinations''} that are \textit{``much too complicated.''} \textbf{P5} stated that they \textit{``have not yet figured out reliably
how to set the timer''} on their stovetop, with \textbf{P4} expressing a similar issue. 
Other participants mentioned successfully using devices without understanding why, attributing it to unintuitive design. For example, \textbf{P3} said \textit{``All I know is I have to push start twice for anything to happen. I have no idea why. I just know that if I push start twice after a few seconds, [my dishwasher] will start.''} Similarly, \textbf{P7} noted that \textit{``it wasn't clear in the beginning that you had to push the [dishwasher's] start button twice,''} with \textbf{P6} following up that \textit{``just the most simple action, turning it on and turning it off, is a mystery.''}

\textbf{\textit{Supporting Superfluous Functionality.}}
Some participants mentioned an overall interest in simplicity, either wanting a more streamlined EID or reducing the amount of technology they use. When discussing dishwashers, \textbf{P10} stated that \textit{``As you age you need simpler models...when you hit 90...we want clean dishes and that's all we care,''} with \textbf{P8} adding that \textit{``Basically, it's too detailed and it should be much simpler.''} \textbf{P9} agreed, noting that \textit{``we don't need all these gimmicks''} with \textbf{P6} further stating that \textit{``We want to know how to turn it on and off.''} \textbf{P11} similarly questioned why the settings for their microwave were not \textit{``just on or off?''}

\subsection{How Do PLA Compensate for Inaccessible EIDs?}
We observed a noticeable amount of participant comments that showed learned methods of working around these usability issues.

\textbf{\textit{Augmenting the EID.}}
To still use the device, participants described three main ways of augmenting the device to increase usability.  
First, we observed additional reminders and instructions placed in close proximity to EID, ranging from small sticky notes to laminated PDF lists of instructions.
Second, some participants mentioned using alternative sensory cues to understand a device's status. For example, \textbf{P4} mentioned that \textit{``when I can't hear [my dishwasher] anymore, I know it's done,''} while \textbf{P5} stated that \textit{``the microwave timer is so faint that I can't even read it, but I can usually smell when it's done.''} \textbf{P10} used the lack of sound to know when their washing machine was finished instead of using a timer or audio signal. 
Third, participants sometimes used entirely different EIDs to perform a specific task. For example, \textbf{P5} mentioned that \textit{``I always set the microwave timer because I can't figure the stove one out.''}, with \textbf{P4} agreeing that the microwave timer was easier to set. 
Multiple participants (\textbf{P1}, \textbf{P3}, \textbf{P4}, and \textbf{P5}) stated that they ignore the information on the digital community information board and instead look it up on their computer or phone, citing reasons of convenience (not leaving their apartment), trusting their own device more than the public display, and preferring the ability to control or select which information is being shown. 

\textbf{\textit{Personalized Interaction Strategies.}}
Participants also adjusted the way they used these devices (if at all) to deal with difficulties in using them. Several participants described consistent habits when using specific devices, often following a short set of steps to achieve a particular outcome.
For example, \textbf{P3} described their dishwasher use as the following: \textit{``I don't change anything because, you know, I just push the start button twice because I do the same thing because I can't read it at all.''} When using their microwave, \textbf{P6} learned that \textit{``basically...if I push one, it'll run for one minute.''} These types of routines also happened with the stepping machine (\textbf{P2} and \textbf{P5}), thermostat (\textbf{P9}), treadmill~(\textbf{P7}), and washing machine (\textbf{P1}). 
Some instead reduced how often they used a device, often choosing to stop using it. For example, \textbf{P7} mentioned never using their dishwasher due to the difficulty reading the labels. Similarly, \textbf{P10} mentioned that they never change their thermostat and \textit{``don't go near''} the digital community information board; \textbf{P9} also mentioned avoiding that device, stating that~\textit{``It bites, gotta keep away.''}

Some mentioned simply ignoring presented information. 
When describing how they use their washing machine, 
\textbf{P4} said that \textit{``I turn it on, I don't pick what temperature it should be. You
know, I pick medium, almost always pick medium, and poof I'm good to go. All the rest of it, oh well!''} \textbf{P1} similarly said their washing machine \textit{``has some lights that go on. But first of all, I can't tell you what they are because I don't ever look at them.''} \textbf{P7} only sets \textit{``the elevation and the speed''} on the treadmill, ignoring other performance metrics and settings because the facility staff \textit{``have told us that it's not accurate.''} 
\textbf{P6} said they \textit{``don't use that little screen at all''} on their microwave, with \textbf{P11} sharing a similar sentiment. \textbf{P10} stated that they do not use the digital community information board because \textit{``it's too much, you know, my life is based on simplifying.''}

\textbf{\textit{Trial and Error.}}
Some participants described a trial-and-error approach to learning to use an EID, with \textbf{P6} saying \textit{``the first time we used [our dishwasher]...it took [my husband] forever...it started and it ran all night...but the dishes weren't dry...''} and \textbf{P7} noting a similar experience. 
\textbf{P10} described finding microwave functions as \textit{``I just try them, and if it doesn't work, I try something else.''}

\section{Discussion}\label{sec:discussion}
In this section, we delve into the major takeaways from our conversations with PLA about their daily encounters with EIDs, including knowledge barriers, accessibility challenges, and possible research questions that could guide future work in this area.

\subsection{Knowledge Gaps and Design Inconsistency Hinder the Accessibility of EIDs for PLA}
Participants' comments indicate that the design of EIDs tends to overlook their probable lack of knowledge and experience when interacting with unfamiliar new displays. To meet the knowledge needs of PLA, embedded information displays should be designed to be self-sufficient, providing the essential information required to learn and operate them independently. However, this can be a challenging and complex task due to the often limited space available for EIDs. Addressing this challenge may require innovative solutions, such as on-demand augmentation of display space using technologies like augmented reality or adaptive displays that dynamically present information according to PLA's needs. 

Participants also noted inconsistent design language and information encoding, such as using unfamiliar symbols and icons, as barriers to transferring and mapping their previously learned knowledge to new EIDs.  This inconsistency can cause uncertainty and hesitation, impeding their ability to fully utilize these features.  Additionally, participants expressed a preference for a specific set of basic functions or a simpler overall design. Incorporating the principles and mandates of universal design into the design of EIDs can probably alleviate issues caused by design variations. Universal design aims to create artefacts that are accessible and usable by the widest range of people, regardless of their abilities or backgrounds. This approach helps ensure that EIDs are intuitive and straightforward for all users, reducing the impact of variations in design and making the interfaces more inclusive and effective.

\subsection{EIDs Complicate Accessibility Considerations}

Compared to visualizations which typically are viewed on electronic screens on personal devices such as phones, computers, and tablets, EIDs present additional challenges for accessible design. Printing text and icons on diverse materials commonly found in home devices (\eDotG stainless steel, wood, marble, plastic) and ensuring they meet contrast standards pose a much wider variety of situations to consider and test during design and production. For PLA, particularly those with visual impairments or in dimly lit environments, small or poorly-legible fonts can notably hinder the ability to read and interpret information. Moreover, electronic screens on devices like thermostats and microwaves often feature limited space, which can lead to interfaces that feel crowded or cluttered, overwhelming PLA trying to locate essential information.
These challenges are compounded by the sheer variety of settings in which EIDs are encountered; it may be above one's head for a thermostat or microwave, or it may be at waist-level for a dishwasher or washing machine. Physical interaction requirements that are awkward or uncomfortable due to factors such as physical location (\eDotG requiring a stool to use a microwave) can further discourage PLA from fully engaging with these displays. EIDs that use an auditory signal should be able to be heard from a reasonable distance away, however they must also avoid disruptions to a home environment by being too loud. Hence, ensuring information accessibility across varied scenarios remains a critical concern in the design of EIDs.

\subsection{Proposed Future Research Questions}
Our observations on EIDs' readability issues and their impact on PLA's device interactions can inform several directions of future work. Many devices in the study such as stoves and microwaves have existed for many years, potentially leading PLA to internalize certain design languages. What assumptions do they have regarding how these devices communicate information, and how do these align with current EID designs? Some participants favored simplicity and an overall reduced usage of technology, which coincides with prior work~\cite{lee2015perspective,vaportzis2017older}. How can EIDs facilitate more streamlined use, and how do they influence decisions to abandon a device? While our study focused on 10 home devices, PLA encounter many others with diverse EID designs. What are the critical dimensions of this design space, and how do they impact usability? A more structured framework could better guide EID design and development.

None of the studied devices connected to personal electronics such as tablets or phones for information sharing or remote control. Existing work in visualization has begun exploring cross-device information design (\eDotG~\cite{brudy2019cross}), which could extend to EIDs in home devices such as displaying wash times on a phone or sending notifications when a task is done~\cite{corno2015context}. What additional considerations arise when sending information between mobile devices and EIDs? This approach might conflict with Wood et al.'s~\cite{wood2007influencing} recommendation to keep EIDs near the monitored device, requiring further investigation. Also, while most EIDs used simple representations, some featured complex visualizations such as a bar graph on the treadmill. When are more detailed visualizations necessary, and would current design guidelines be inclusive toward PLA? Addressing these questions could lead to more informative, user-friendly, and accessible EIDs.

\section{Conclusion}
We examined PLA's experiences using embedded information displays in home devices.  Conducting two focus groups at a nearby retirement community, we discussed EIDs from a set of common everyday home appliances and electronics, focusing on improvement areas for designing how these communicate information to a user. We observed major themes regarding difficulties for PLA while interacting with the device, problems with reading information from the device, and compensating for these difficulties in various ways. We discussed how these observations could lead to improvements in the design of future devices and incorporate existing accessibility knowledge, looking ahead to a future with more usable home electronics for this growing population.

\acknowledgments{
Tanja Blascheck is funded by the European Social Fund and the Ministry of Science, Research and Arts Baden-Württemberg.}

\balance
\def\UrlBreaks{\do\/\do-}
\bibliographystyle{abbrv-doi}
\bibliography{bibliography.bib}

\end{document}